\documentclass[fleqn,usenatbib]{mnras}

\usepackage[T1]{fontenc}

\DeclareRobustCommand{\VAN}[3]{#2}
\let\VANthebibliography\thebibliography
\def\thebibliography{\DeclareRobustCommand{\VAN}[3]{##3}\VANthebibliography}

\usepackage{graphicx}	
\usepackage{amsmath}	
\usepackage{amssymb}	

\usepackage{color}
\usepackage[dvipsnames]{xcolor}

\DeclareGraphicsExtensions{.png,.pdf,.eps,.jpg,.tiff,.ps}

\newcommand{\vunit}{\mbox{\,km\,s$^{-1}$}}

\newcommand{\Lsun}{\mbox{\,$L_\odot$}}

\newcommand{\mic}{\mbox{$\,\mu$m}}

\newcommand{\ltsimeq}{\raisebox{-0.6ex}{$\,\stackrel 
	{\raisebox{-.2ex}{$\textstyle <$}}{\sim}\,$}}

\usepackage{graphicx}	
\usepackage{amsmath}	
\usepackage{amssymb}	

\definecolor{cyan(process)}{rgb}{0.0, 0.72, 0.92}

\newcommand{\commt}[1]{\textcolor{cyan(process)}{#1}}

\title[The recurrent nova M31N 2008-12a]
{Near-Infrared Spectroscopy of the Recurrent Nova M31N 2008-12a}
\author[D. P. K. Banerjee et al.]{D. P. K. Banerjee,$^{1}$\thanks{E-mail: dpkb12345@gmail.com}
T. R. Geballe,$^{2}$ A. Evans,$^{3}$ C. E. Woodward,$^{4}$
K. L. Page,$^{5}$ \newauthor
S. Starrfield$^{6}$ \\
$^{1}$Physical Research Laboratory, Navrangpura,  Ahmedabad, Gujarat 
380009, India\\
$^{2}$Gemini Observatory/NSF's NOIRLab, 670 N. Aohoku Place, Hilo, HI, 96720, USA \\
$^{3}$Astrophysics Group, Lennard Jones Laboratory, Keele University, Keele, Staffordshire,  ST5 5BG, UK\\
$^{4}$Minnesota Institute for Astrophysics, School of Physics \& Astronomy, 116 Church Street SE, University of Minnesota, \\
Minneapolis, MN 55455, USA \\
$^{5}$School of Physics and Astronomy, University of Leicester, University Road, Leicester, LE1 7RH, UK \\ 
$^{6}$School of Earth and Space Exploration, Arizona State University, 
Box 876004, Tempe, AZ 85287-6004, USA\\
} 

\date{Accepted XXX. Received YYY; in original form ZZZ}

\pubyear{2025}

\begin{document}
\label{firstpage}
\pagerange{\pageref{firstpage}--\pageref{lastpage}}
\maketitle

\begin{abstract}
Near infrared (NIR) 0.9-2.5\mic\ spectra of 
the remarkable recurrent nova M31N 2008-12a were
obtained on days 6.3 and 10.3 after discovery
of its 2024 outburst, and are the first NIR spectra of this
object. The only prominent line seen in
the spectra is that of \ion{He}{i} 1.083\mic, on day 6.3. Apart 
from this \ion{He}{i} line, 
there are only two other weak emission 
features: one at 1.0786\mic, suggested to be the 
[\ion{Fe}{XIII}] 1.075\mic\ coronal line,
and one unidentified feature at 1.0969\mic.
The observed full width at half maximum of the \ion{He}{I}
line on day~6.3 (1350\vunit) 
is consistent with the behaviour of
optical \ion{He}{i} lines during earlier eruptions of this
RN, which show that the nova ejecta decelerate as they
interact with the secondary's wind.
The \ion{He}{I} 1.083\mic\ line faded rapidly, and 
was absent in the day 10.3 spectrum, along with any other
emission lines. 
We use the relative strengths of optical He and H 
lines in previous eruptions to estimate the expected 
strengths of the \ion{He}{I} 1.083\mic\ line and of other 
infrared (including coronal) lines at 6.3~days after eruption.
Our findings are consistent with the infrared spectra we 
observed during the 2024 eruption. We
apply our analysis to account for the relative weakness of
NIR coronal emission in known Galactic recurrent novae
with giant secondaries.
\end{abstract}




\section{Introduction}

Nova explosions occur on the surfaces of white dwarf
(WD) stars in semi-detached binary systems. The other
star in the binary, the secondary, may be a main sequence
star, a sub-giant or a red giant (RG). Material from the
secondary flows onto the surface of the WD via the inner
Lagrangian point, and forms an accretion disc 
\citep[see][for a review]{bode12}. 
Eventually conditions at the base of the mixed layer
become suitable for a thermonuclear runaway (TNR), which
results in the explosive ejection of material. As a
result of the TNR, the ejected material is enriched in
metals, up to Ca \citep*{he24,starrfield24}. After the
nova eruption has subsided, mass transfer from the
secondary resumes and, in time, the above process is
repeated. All novae must eventually recur, but 
{\em recurrent novae} (RNe) do so
on timescales $\ltsimeq100$~yrs.

If the RN has a RG secondary, the material ejected
in the eruption collides with, and shocks, the RG wind,
and a reverse shock is driven into the ejecta. 
This process may also occur if the secondary
is a sub-giant \citep[as in U~Sco;][]{evans23}, 
provided the mass ejected in the eruption is low.
This results in $\gamma$-ray, X-ray and non-thermal
radio emission, and
coronal line emission in the ultraviolet, optical and
infrared  \citep[IR; see][]{dellavalle20}.

Extragalactic RNe are particularly useful for
understanding the RN phenomenon because (unlike
Galactic RNe) their distances are accurately known. Fewer
than a dozen Galactic RNe are known, but 
twenty two RNe have been identified in M31
\citep*{healy24,shafter25}. Of these, the
remarkable system M31N 2008-12a (hereafter M31-12a) has
erupted annually since its initial discovery. Its most
recent eruption occurred on 2024 December 12.
We present here near-infrared (NIR)
spectroscopy of the 2024 December eruption.

\section{M31-12a}

The first known eruption of M31-12a was discovered in 1992
January \citep{henze18}, and it is known to have had annual
eruptions since 2008 \citep{darnley20,darnley21}. 
Its inter-eruption period is
estimated to be $0.99\pm0.02$~yr \citep{darnley21}, the
shortest of any known RN. 
The nature of the secondary star in M31-12a has not
been determined with any certainty, but \cite{darnley15}
consider a RG is the most likely
\citep[see also][]{williams16}. This would be consistent 
with \cite{henze18}'s estimate of its orbital
period ($350{\la}P_{\rm orb} \mbox{~(in days)~}\la490$,
depending on the secondary mass). The lower orbital period 
estimate by \cite{darnley17b} 
($5{\la}P_{\rm orb} \mbox{~(in days)~}\la23$) would
also allow a RG classification. We shall assume a
RG secondary here.

Comprehensive studies from
2013-2022 \citep{darnley16, basu24} suggest that the time
$t_3$ for the light curve to decay by three magnitudes from 
maximum is wavelength-dependent, but extremely short, e.g.
\commt{$\ltsimeq2.5$~days in the $V$} band.
M31-12a's faintness and rapid decay make spectroscopic 
observations very difficult. The earliest (optical)
spectra were obtained within hours of the peak of the
2014 eruption \citep{darnley15}. Most optical spectra have
been obtained within six days of maximum 
\citep{darnley17a, henze18}. 
The latest optical spectrum seems to have been obtained 
$\simeq18.8$~days after maximum \citep{darnley17b}, and 
showed a continuum and no emission lines. \cite{shafter11}
carried out a photometric and spectroscopic survey
of M31 novae with the Spitzer Space Telescope
\citep{werner04}, but there are no 
reported NIR spectra of M31 novae. The spectra 
presented here are  among the latest\footnote{``latest''
in the sense of time elapsed since outburst.}
spectra taken of M31-12a, and only
the second NIR spectra of any extragalactic nova  
\citep[the first being that of the LMC RN LMCN 
1968-12a,][]{evans25}. 

The heliocentic radial velocity of M31 is $-301$\vunit\
\citep{courteau99}, but in the case of M31-12a, this is
largely offset by the rotation of M31 \citep{zhang24},
with radial velocity $+200$\vunit,
with a likely range of $\pm25$\vunit. 
We take the effective heliocentric radial velocity of 
M31-132a to be $-80$\vunit. We assume a distance 770~kpc
and reddening $E(B-V)=0.1$ \citep{darnley17b}. The 2024
December eruption of M31-12a was reported by 
\cite{zhao24a}, 
with an earlier detection on 2024 December
12.9875~UTC \citep{zhao24b}. We take the latter
as our time origin, $t_0$.

\section{Observations}
\subsection{Near infrared observations}\label{NIR}
NIR (0.8--2.5\mic) spectra of M31-12a were obtained
at the Frederick C. Gillett Gemini North Telescope on 
Maunakea on UT 2024 December 19 between 06.95 and 08.17 hrs UT,
and on December 23 between 07.04 and  08.57 hrs UT. 
The facility instrument GNIRS \citep{elias06}
was used in its normal cross-dispersed mode using the 
$32\ell$/mm grating and the 1\farcs0 wide slit aligned
along the mean parallactic angle on each night. 

GNIRS acquisition is normally done in the $H$-band.
However, M31-12a was anticipated to be faint,
hence blind-offsets were performed from
the $H=13.3$ star Gaia DR2 375301565245820416, 52\farcs8
from the nova. The off-setting has an accuracy of better
than 0\farcs1.
The data were obtained in the standard ABBA nod-along-slit
mode. The total exposure time was
4,800~s on the first night, and 7,200~s on the second. 
Data reduction used a combination of IRAF \citep{tody86}
and Starlink Figaro \citep{curry14} as described, for example, in
\cite{evans22a}. The final flux-calibrated
spectra presented here are binned in 0.001\mic\
segments in the 0.82--1.50\mic\ intervals and in 0.002\mic\
bins in the 1.50--2.52\mic\ intervals. 
The wavelength calibration, determined from
concurrent observations of an argon arc lamp, is accurate to 
better than 0.001\mic\ ($\la280$\vunit\ at 1.083\mic).

\subsection{X-ray observations}

{\it Swift} began daily monitoring of the nova at 20.5 UT
on 2024 December 13 ($t_0$+0.87 d). An ultraviolet
source with $uvw1 = 17.23\pm0.05$ was promptly detected,
although no X-ray
emission was seen until December 18 ($t_0$+5.52 d), when a 
super-soft source was seen \citep{page24a,page24b}. 
The {\it Swift} data were analysed using {\sc heasoft}
v6.34, together with the relevant calibration files.

\section{Results}
\subsection{The \ion{He}{i} 1.083\mic\ emission line}

\begin{figure*}
 \includegraphics[width=10.5cm,keepaspectratio]{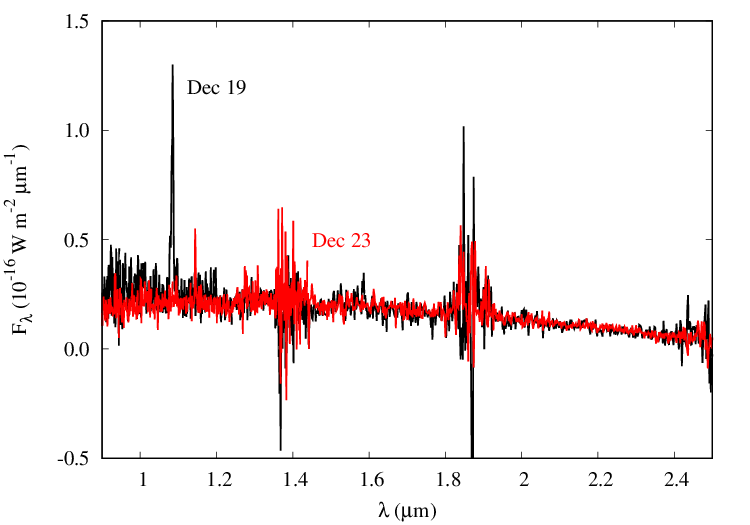}
 
 \vspace{5mm}
 
  \includegraphics[width=10.5cm,keepaspectratio]{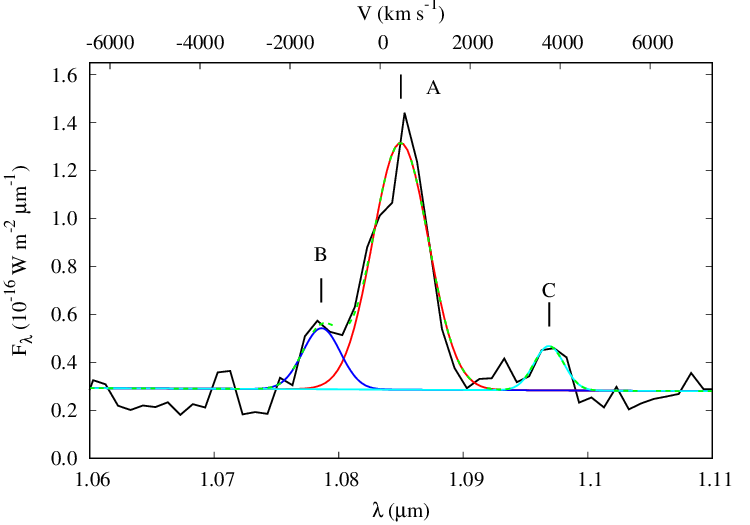}
 \caption{Top: observed spectra (undereddened) on days 6.3
 (2024 December 19) and 10.3 (2024 December 23). The noisy
 regions around 1.4\mic\ and 1.85\mic\ are due to poor 
 atmospheric transmission. 
 Bottom: Spectrum in the region of the \ion{He}{i} triplet, showing 
 the main peak (A) and two weak lines (B and C) in the wings.
 Data dereddeded by $E(B-V)=0.1$, and shifted to the rest
 frame of M31-12a. Gaussian fits are shown to the
three components.
 Black line: data, red line, gaussian fit to A,
 blue line, gaussian fit to B, cyan, gaussian fit to C.
 Green line is total fit.
 The velocity scale is for a rest wavelength corresponding to
 the 1.083\mic\ line in the reference frame of M31-12a.  See text for discussion.
 \label{All_data}}
\end{figure*}

The spectra are shown in Fig.~\ref{All_data}. The only prominent
emission line seen is the \ion{He}{I} 1.0833\mic\ 
triplet (hereafter the 1.083\mic\ line;
note that the long and short wavelength components of the triplet are separated by only 0.000125\mic)
on day~6.3, when the peak-to-continuum ratio is
about 7; the other line parameters are given in 
Table~\ref{fluxes}. 
This feature appears to have three components 
(labelled A--C in Fig.~\ref{All_data}).
The centroid of the  main peak (A) was 1.0845\mic; 
its deconvolved FWHM was $\sim1350$\vunit. As the
accuracy of the wavelength calibration is better than 280\vunit,
the shift (+330\vunit) of the observed line from
the rest wavelength of the \ion{He}{i} line appears real. 
The weaker features (B and C) are discussed in 
Section~\ref{coronals} below.

\begin{figure}
\includegraphics[width=7.5cm,keepaspectratio,angle=0]{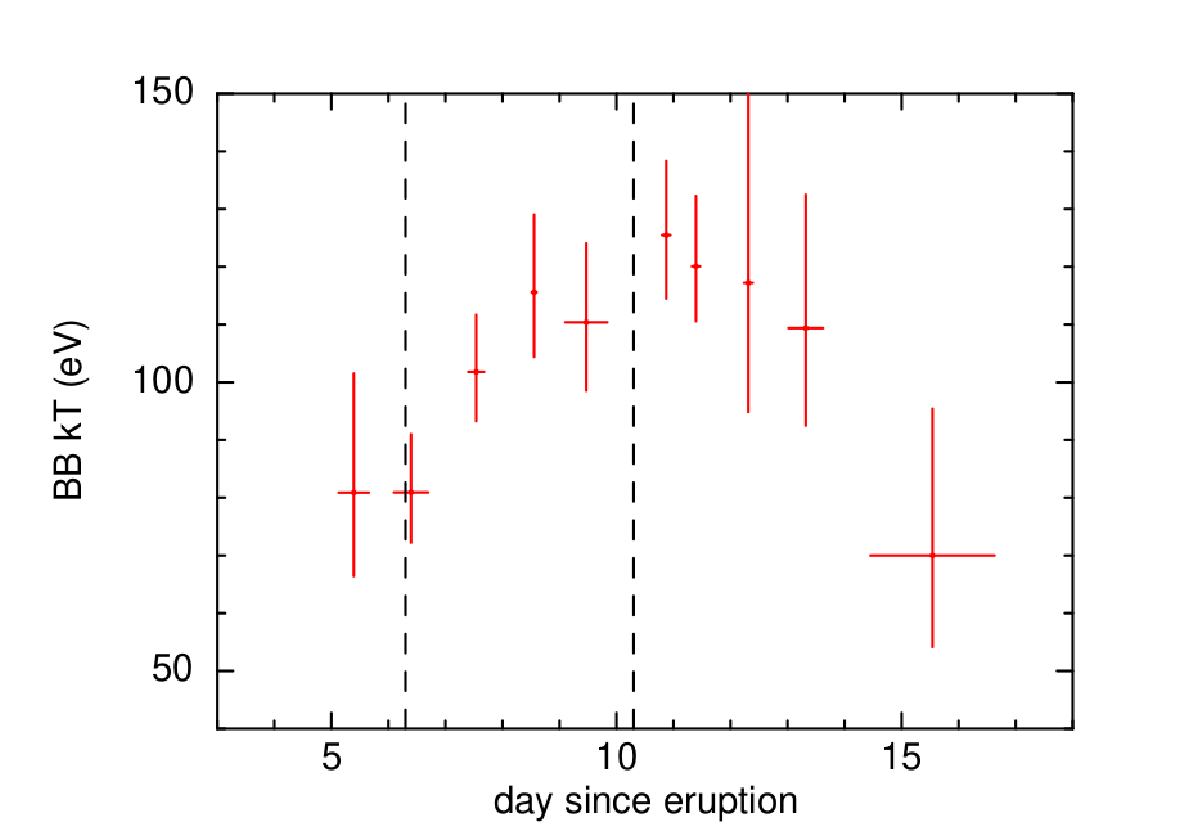}
\caption{Temperature of the super soft X-ray source as determined
by fitting a blackbody to {\it Swift} data. The dashed lines
indicate the dates of our observations.\label{swift}}
\end{figure}

No emission lines were detected
on day~10.3, the 1.083\mic\ line having
faded below the $3\sigma$
detection limit of $10^{-17}$~W~m$^{-2}$\mic$^{-1}$. 
The reason for its disappearance could be a combination
of two reasons: (i)~the recombination line intensity is dependent
on $n_en(\mbox{He}^+$). Dilution of the ejecta because of
expansion will reduce both $n_e$ and $n_i$ as 
$\propto{r}^{-2}$. Hence $n_en_i$ will fall by a
factor of $(6.3/10.3)^4\simeq0.14$ between days 6.3 and 10.3;
and (ii)~a possible shift in the ionisation equilibrium of He 
resulting in a larger fraction of He$^{+2}$ in preference to
He$^{+1}$, due to the hardening radiation from the central WD. 
The evolution of the radiation field of
the central WD with time is shown in Fig.~\ref{swift},
where the model WD black body (BB) temperatures, based on
{\it Swift} data, are shown. The BB temperatures are approximately 
$8.8\times10^5$~K and $1.27\times10^6$~K on days 6.3 and 
10.3 respectively.

\begin{figure}
\includegraphics[width=7.5cm,keepaspectratio]{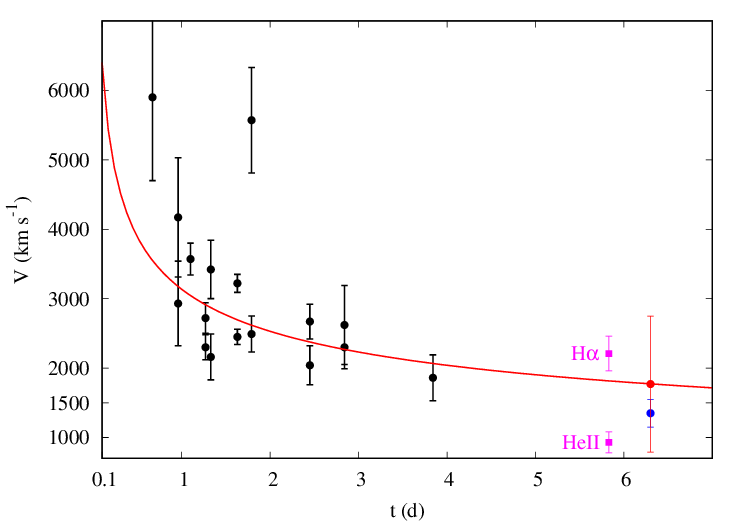}
\caption{Fit of power-law to FWHM of the optical \ion{He}{i} lines
from 2014-15 data. Blue point at 6.3~days is from the data
reported here. Red point at 6.3~days is extraolation of fits, with 
uncertainties from uncertainties in $A$ and $\alpha$. 
Magenta points at 5.83~days are H$\alpha$ and \ion{He}{ii}
4686\AA\ data from \protect\cite{henze18}. \label{fits}}
\end{figure}

The $^3$P$^o-^3$S \ion{He}{i}
transition at 7065\AA, and the $^3$P$^o-^3$D transition at 5876\AA,
both end up at the upper level of the 1.083\mic\ line.
It seems reasonable therefore to expect that the FWHMs of the
5876/7065\AA/1.0833\mic\ lines behave in the same way.
We have taken FWHM data for the 5876\AA\ and 7065\AA\ lines
for the 2014-15 eruptions from Table 6 of 
\cite{darnley16}, and fitted a function of the form 
$V=A\,t^{-\alpha}$, giving (with $V$ in \vunit, $t$ in days
from outburst; see Fig.~\ref{fits})
\begin{equation}
 V   =  [3135\pm285] \:\: t^{-0.31[\pm0.17]} ~~. \label{decel}
\end{equation}
The exponent $\alpha\simeq\frac{1}{3}$
is similar to those fitted to the variation of the H$\alpha$
FWHM by \cite{darnley15}. \cite{darnley16} find
$V\propto{t}^{-0.28\pm0.05}$ for H$\alpha$ (2015 eruption), while
\cite{basu24} report $V\propto{t}^{-0.27\pm0.07}$ from a mix
of eruptions and lines (H, He). The value of
$\alpha=\frac{1}{3}$ corresponds to ``Phase~II'' of the
development of the remnant, when the outer shock has radius
$r_{\rm shock}\propto{t}^{2/3}$ \citep{bode85}.

The ejecta are decelerated by the secondary wind,
and when they reach the edge of the wind we have ``break out''.
For wind  speed $w$, ejecta speed $V$, time since last
eruption $T$, time to break out is $t_{\rm b-o}\sim{wT}/V$,
but this does not take into account the deceleration of the ejecta.
Integrating $At^{-\alpha}$ in Equation~(\ref{decel}) 
gives the time to break out as 
\begin{equation}
 t_{\rm b-o}  =  \left [ \frac{wT(1-\alpha)}{A}
     \right ] ^{1/(1-\alpha)}  ~~.  \label{b-o}
\end{equation}
RG wind velocities can be in the range 10--45\vunit\ 
\citep*{wood16,wood24}. Taking the ejecta speed to be 2950\vunit\
(half the first FWHM value, 5900\vunit), we find
\begin{eqnarray}
 t_{\rm b-o} & \simeq 2.0 & \left ( \frac{w}{20\vunit} \right )^{1.45}
\left ( \frac{V}{2950\vunit} \right )^{-1.45} \nonumber \\
 && \times \left ( \frac{T}{\mbox{1~year}} \right )^{1.45} \mbox{~~days.} 
\end{eqnarray}
This suggests that the ejecta might have reached the edge of
the RG wind within a couple of days of outburst, but 
$t_{\rm b-o}$ could be as large as 5.4~days if $w\simeq40$\vunit.
Equation~(\ref{decel}), if extrapolated, suggests
$V=1770\pm980$\vunit\ at $t=6.3$~days. The observed FWHM on
day 6.3 is 1350\vunit, so extrapolation of the implied
deceleration suggests a value consistent with the observed value.
This indicates that deceleration might indeed 
have continued until this later time.

Further support for this possibility is provided by the
data of \cite{henze18}, who obtained an optical spectrum 
5.83~days (close to our 6.3~days) after the 2016 eruption.
They detected \ion{He}{ii} 4686\AA, usually taken to 
be an accretion disc signature, with FWHM $930\pm150$\vunit.
On the other hand, their FWHM for H$\alpha$, $2210\pm250$\vunit\
on day 5.83, is close to our fitted curve (see Fig.~\ref{fits}). 
This, together with the closeness of the extrapolated deceleration,
might suggest that deceleration continued until at least day~6.3.
Observations up to $\sim$~day~6 are necessary to 
examine this possibility.

\begin{table*}
\caption{Line fluxes for December 19, dereddended by
$E(B-V)=0.1$, and shifted to rest frame of M31-12a.
Line identifications correspond to the labelling in
the bottom panel of Fig.~\ref{All_data}. \label{fluxes}}
 \begin{tabular}{ccccc}
  Line centre & Fig.~\ref{All_data} & ID & FWHM & Line flux \\ 
  (\mic)  & label &  & (\vunit) & ($10^{-20}$~W~m$^{-2}$) \\ \hline
  1.0850  & A & \ion{He}{i} 1.0833\mic & 1350  & $57.6\pm2.6$ \\
  1.0786  & B & See Section~\ref{fexiii} & 640 & $9.7\pm0.9$ \\
  1.0969  & C & Unidentified & 540  & $5.6\pm0.9$ \\
  \hline\hline
 \end{tabular}
\end{table*}

\subsection{The lack of hydrogen lines}\label{lack}
We have explored the reason for the absence of H lines and
other He lines, in the  day 6.3 spectrum. Several Paschen and
Brackett series lines are routinely seen in the NIR spectra of
classical novae and RNe.
We used the observed H and He line fluxes from 
\cite{darnley16} (their Table 7), and derived mean values
(averaged over all epochs) for the ratio of the H$\alpha$
intensity relative to the He lines 5876\AA, 6678\AA\ and 7065\AA.
The predicted He line intensities from \cite{porter12} and 
\cite*{benjamin99}, for a reasonable range of electron 
densities ($10^6-10^9$~cm$^{-3}$) likely to exist in the 
ejecta on day 6.3, and for an electron temperature 
$T_e = 10^4$~K, were used to estimate the ratios of 
1.083\mic\ relative to the above 
three optical He lines. These ratios were then used to
estimate the relative strength of \ion{He}{I} 1.083\mic\ 
to H$\alpha$. The expected \ion{He}{I}(1.083)/H$\alpha$ ratio is
found to be $6.3\pm2.4$, showing that the 1.083\mic\ line can be
strong. The 1.083\mic\ line can be dominant in nova spectra
during the late nebular phase, greatly surpassing the strengths
of other H and He lines in the NIR region 
\citep[several examples are given in][]{banerjee18}. 

\begin{figure}
 \includegraphics[width=7.5cm,keepaspectratio]{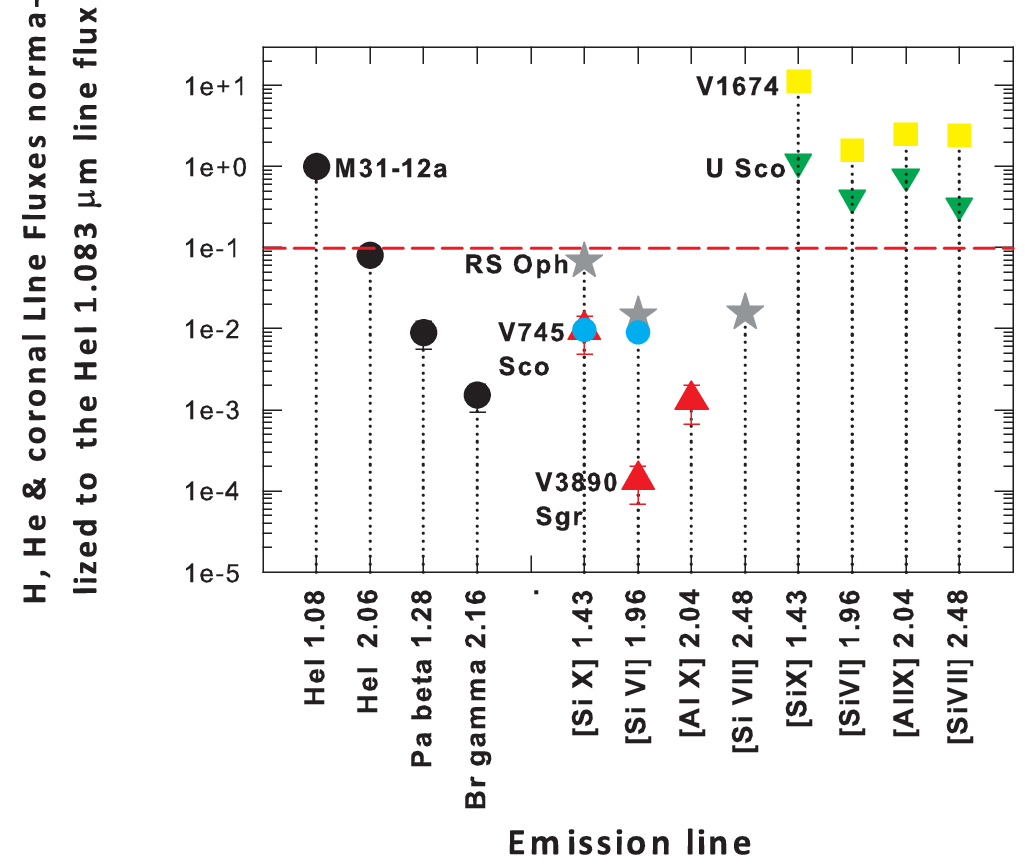}
 \caption{Black circles: expected strengths in M31-12a
 of three of the strongest H and He NIR recombination lines
 relative to \ion{He}{I} 1.083\mic\ (set to
 unity). Dashed red line: flux level at 10\%
 of the \ion{He}{I} line. Grey stars, red triangles and 
 cyan circles are observed strengths of some of the
 strongest coronal lines normalised to \ion{He}{I}
 1.083\mic\ for RS Oph, V3890 Sgr and V745 Sco. Coronal
 line fluxes were measured from spectra as close as 
 possible to the peak of the super-soft X-ray emission
 when coronal line emission is expected to be present
 (RS Oph \citep[day 55.7;][]{evans07}, V3890 Sgr
 \citep[day 15.1;][]{evans22b}, and V745 Sco
 \citep[day 8.7;][]{banerjee14}). 
 Yellow squares and green upturned
 triangles are for two representative novae (U Sco and
 V1674 Her) without a RG donor. U Sco is on day 28.4
 from \protect\cite{evans23}, and V1674 Her is on day 50.4 from
 \protect\cite{woodward21}.
 \label{lines}}
\end{figure}

Based on the expected \ion{He}{I}~1.083/H$\alpha$ ratio,
the strengths of other H and He lines under 
Case~B conditions were calculated from \cite{storey95}
and \cite{porter12}. In Fig.~\ref{lines} are plotted the H and
He extinction-corrected flux ratios using the best
available $E(B-V)$ for some of the strongest lines in
the NIR region, namely Pa$\beta$ 1.2816\mic, Br$\gamma$
2.1661\mic, and \ion{He}{I} 2.0587\mic\  (black circles). 
It can be seen that these lines are weak ($\la0.1$ 
times that of the 1.083\mic\ line). Thus they are not 
expected to be seen in our NIR data given the 
signal-to-noise level on  the continuum.

\subsection{Coronal lines}
\label{coronals}
\subsubsection{The [Fe XIII] 1.0750\mic\ coronal line}
\label{fexiii}
A motivation for this study was to try and detect
coronal line emission from elements such as  Si, S, Ca, Mg,
Al, and P, which are often present during the coronal phase 
\citep*[e.g.,][]{banerjee09,evans22a,evans22b},
together with lines of the Pa and Br series.
A comparison of the observed strengths of coronal lines
with model nucleosynthesis predictions can tell us about
the mass and nature of the WD (CO or ONe). 
Furthermore, analysis of the H recombination
lines will throw light on optical depth effects
in the ejecta, and allow us to place constraints on the
ejected mass.

Elemental yields of the above elements can be
vastly enhanced in nova explosions, yet be significantly different
between ONe and CO novae. Even within one class (e.g. the ONe
novae), the elemental abundances can differ widely with the
mass of the WD \citep{starrfield24,starrfield25}. Thus, 
coronal line analysis can give important insights about the WD. 

On the blue wing of the 1.083\mic\ line there
is a feature
(labelled B in the bottom frame of
Fig.~\ref{All_data})
centred at 1.0786\mic, which we tentatively
identify as the
[\ion{Fe}{XIII}] 1.0750\mic\ coronal line. No line is
expected at this wavelength in
\ion{Fe}{II} or He/N novae during the early stages
\citep[see, e.g.,][]{banerjee12}. 
The difficulty with the [\ion{Fe}{XIII}] assignment is that the 
line is redshifted by $\sim1000$\vunit\ from its rest wavelength
(see Table~\ref{fluxes}). However, the wavelength mismatch is
mitigated somewhat by noting that the 1.083\mic\
line is also seen to be red-shifted, although by a smaller
amount, $\sim330$\vunit. 
These shifts are greater than can be attributed to
uncertainty in the wavelength calibration 
(see Section~\ref{NIR}).
The reason(s) for the redshift(s) is
unclear. There is another weak (but real) feature on the red
wing, at the wavelength of Pa$\gamma$ 1.094\mic\
(labelled C in Fig.~\ref{All_data}).
However this can not be Pa$\gamma$ as the
stronger Pa$\beta$ 1.2818\mic\ is not seen
(see Section~\ref{lack}). 
This feature remains unidentified.

As an alternative to
the [\ion{Fe}{XIII}] assignment, we 
consider whether the 1.0786\mic\ feature, together with the
1.094\mic\ feature, could constitute a pair of blue and red high
velocity components of \ion{He}{I}, as might be
expected in M31-12a.
High velocity material was seen 
in the 2015 eruption by \cite{darnley16}
in the line profiles of H$\alpha$,
H$\beta$ as well as in those of 
\ion{He}{I} 5876\AA\ and 7065\AA. The high
velocity flow was seen as a broad, rectangular pedestal,
symmetrically placed below the brighter central emission feature,
with a large FWZI in the range 12000--14500\vunit.
However, this high velocity flow was short-lived; it was
seen on days 0.67, 0.96 and 1.1, and faded over this period. 
After day 1.1, any emission from such high-velocity material
was either absent, or at least indistinguishable from the
continuum \citep{darnley16}. 
If the 1.097\mic\ feature is redshifted \ion{He}{i},
the implied velocity is $\sim3800$\vunit. Furthermore, the
observed 1.078\mic\ and 1.094\mic\ lines are not symmetric
about the central 1.083\mic\ component, nor are they 
symmetric about the observed centroid of the line, 1.0845\mic.

Further, as 
discussed above, any high velocity component should 
have faded below detection limit by day~6.3.
Thus, the [\ion{Fe}{XIII}] assignment may be correct. 
This is supported by optical spectroscopy of
previous outbursts of M31-12a.
\citep{darnley16} combined spectra, covering 
0.67 to 3.84 days post-eruption, from the 2014 and 
2015 eruptions, and tentatively identified several Fe
coronal lines (e.g., [\ion{Fe}{X}] 6375\AA,
[\ion{Fe}{XIV}] 5303\AA).

The [\ion{Fe}{XIII}] 1.0750\mic\ line is not generally seen
in the NIR spectra of novae. It may be present, but weak,
and thus goes undetected because of the stronger emission from the
blue wing of the 1.083\mic\ line. However, this line
was weakly but clearly detected in the coronal spectra of 
V3890~Sgr \citep{evans22b}, RS Oph (Woodward et al.,
in preparation) and V745~Sco \citep[re-analysis of the data
in][]{banerjee14}. These three RNe, along with T~CrB,
constitute the four known Galactic RNe with RG secondaries
with a wind. It may therefore be no
coincidence that this line is seen in M31-12a,
which may also be a system with a RG secondary.
However, our data do not enable us to
rule out that the wing features are due to \ion{He}{I}.
If the blue feature is due to [\ion{Fe}{XIII}], we 
suggest that it arises from the shock heated wind of the RG.
{\em A spectrum at a higher signal-to-noise during the earliest
($\la{t}_0+3$~days) phase is required to resolve this issue.}

\subsubsection{The absence of other coronal lines}
Fig.~\ref{lines} explores the observed  strength of  
coronal lines in novae with  respect to that of the
1.083\mic\ line. Two classes of novae
are shown: those with a RG secondary
\citep[V745 Sco, RS Oph, V3890 Sgr, hereafter  
``T~CrB systems'';][]{warner12}, 
and those with evolved (but not RG) secondaries,
for example, V1674 Her and U~Sco. The strongest
NIR coronal lines are chosen as benchmarks, viz.,
[\ion{Si}{VI}] 1.96\mic, [\ion{Si}{x}] 1.43\mic, 
[\ion{Al}{ix}] 2.04\mic, and [\ion{Si}{VII}] 2.48\mic. 
Noting the logarithmic scale, 
it is clear  that coronal  emission in these 
lines is weak in T~CrB systems.
The fact that M31-12a is a similar system might
explain why none of the  conventional strong coronal 
lines were seen in its spectrum on day~6.3. 
On the other hand, both U~Sco and V1674 Her show coronal
emission at a strength comparable to, and sometimes
surpassing, that of \ion{He}{I}. 

\subsection{The continuum}
\begin{figure}
  \includegraphics[width=7.5cm,keepaspectratio]{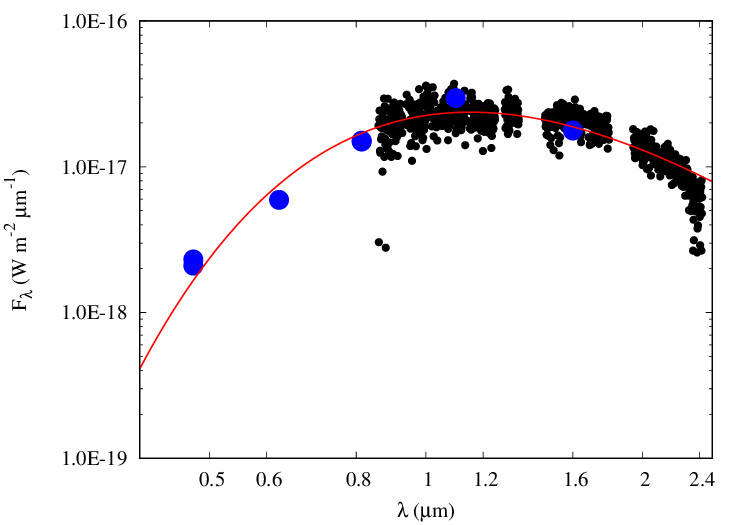}
 \caption{Blackbody fit to the day 10.3 spectrum, 
 to understand the origin of the continuum, which is not that
 of M31-12a (see text).
 \label{continuum}}
\end{figure}
The continuum seen in the spectrum of day 10.3 is unlikely
to be that of  the nova. Rather  there is a strong 
contribution from a nearby star (henceforth Star~1) 
separated from the nova by 0\farcm0084 = 0\farcs5,
with the ``MatchID'' number 102998460 in the Hubble 
Source Catlog HSCV3. IR images of the field around
this star can be seen in the 1.1\mic\ and 1.6\mic\
HST images in the The Panchromatic Hubble Andromeda 
Treasury program \citep{dalcanton12}. 

This star is also shown as Star 1 in Figure~3 of \cite{darnley14}.  
During our observations, the nova and Star 1 happened to be
positioned across the slit, hence along the same dispersion
axis so their spectra largely overlapped. We believe
such an overlap occurred because good agreement is
obtained when the observed day~10.3 continuum is 
matched with the fluxes of this star, as shown in
Fig.~\ref{continuum}. 
In Fig.~\ref{continuum}, the 
fluxes of Star~1 were derived from its AB magnitudes
in the HSCV3:  24.06 (F336W filter), 26.27 (F475W),
24.44 (F625W),  22.77 (F814W),
21.31 (F110W), and  21.01 (F160W). 
In contrast, the day 6.3 continuum clearly
rises above the day 10.3 continuum at wavelengths 
$\ltsimeq1.1$\mic. Thus,  the continuum
below 1.1\mic\ does indeed contain a significant
contribution from the nova itself.
A BB fit to the Star~1 fluxes 
indicates it is a RG with a temperature $\sim2400$~K
and luminosity $L\sim835$\Lsun. 
In view of the above, we can extract no information 
from the day~10.3 continuum
as it is not that of the nova.

\section{Conclusions}
\begin{enumerate}
 \item NIR spectra of the 2024 outburst of the RN M31N 2008-12a
 were obtained on days 6.3 and 10.3 after discovery, the
 first NIR spectra of this RN. 
\item The only prominent line seen in the day 6.3 spectrum is 
\ion{He}{I} 1.083\mic. There is one other weak emission feature
at 1.078\mic, which we suggest is the [\ion{Fe}{XIII}]
1.075\mic\ coronal line. 
\item The observed deconvolved FWHM of the \ion{He}{I} line,
1350\vunit, on day~6.3 is consistent with earlier optical results 
\citep[e.g.,][]{darnley16} that the nova ejecta decelerate as they
interact with the secondary wind.
\item The intensity of the \ion{He}{I} 1.083\mic\ line
faded rapidly: the day 10.3 spectrum is devoid of emission
lines, and shows only a continuum
from a field star. Future NIR observations of this RN
can only usefully be made if they are obtained within the first 
5--6 days after outburst.
\end{enumerate}

\vspace{-5mm}

\section*{Acknowledgements}

The Gemini observations were made possible by award of
Director's Discretionary Time for programme
GN-2024B-DD-106. 

We thank the referee for their helpful comments
that helped to improve this paper. We also thank
Matt Darnley for providing useful information about
his extensive spectroscopic observations of this target.

The international Gemini Observatory is a program of 
NSF's NOIRLab, which is managed by the Association of Universities 
for Research in Astronomy (AURA) under a cooperative agreement 
with the National Science Foundation, on behalf of the Gemini 
Observatory partnership: the National Science Foundation 
(United States), National Research Council (Canada), 
Agencia Nacional de Investigaci\'{o}n y Desarrollo (Chile), 
Ministerio de Ciencia, Tecnolog\'{i}a e Innovaci\'{o}n (Argentina), 
Minist\'{e}rio da Ci\^{e}ncia, Tecnologia, 
Inova\c{c}\~{o}es e Comunica\c{c}\~{o}es (Brazil), 
and Korea Astronomy and Space Science Institute (Republic of Korea).

KLP acknowledges funding from the UK Space Agency.
SS acknowledges partial support from an ASU Regents' Professorship.

\vspace{-5mm}

\section*{Data availability}
The raw infrared data are available from the Gemini Observatory
Archive, https://archive.gemini.edu/~.%
The Swift data are available from
https://www.swift.ac.uk/swift\_live/


\bsp	
\label{lastpage}
\end{document}